\documentclass[11pt,twoside]{article}
\usepackage{asp2010}

\resetcounters

\newcommand{\logg}{\mbox{$\log g$}}
\newcommand{\loggw}[1]{\mbox{$\log g\hspace{-0.5mm} =\hspace{-0.5mm}  #1$}}
\newcommand{\sla}{\raisebox{-0.10em}{$\stackrel{<}{{\mbox{\tiny $\sim$}}}$}}
\newcommand{\Teff}{\mbox{$T_\mathrm{eff}$}}
\newcommand{\Teffw}[1]{\mbox{$\Teff\hspace{-0.5mm}=\hspace{-0.5mm}#1\,\mathrm{K}$}}

\begin{document}

\title{Quality Control for Theoretical Data in the Virtual Observatory:\\
       Establishing Benchmark Tests for Synthetic Spectra}
\author{Thomas Rauch and Ellen Ringat}
\affil{Institute for Astronomy and Astrophysics,
       Kepler Center for Astro and Particle Physics,
       Eberhard Karls University,
       Sand 1,
       D-72076 T\"ubingen,
       Germany\\
       email: gavo@listserv.uni-tuebingen.de}

\begin{abstract}
The Virtual Observatory (\emph{VO}) provides access to both, data and theory. 
Quality control is a general problem and the \emph{VO} user needs partly some 
experience to judge the reliability of the \emph{VO} products. As far as spectral 
analysis is concerned, many different areas are involved, from atomic data 
to stellar model-atmosphere codes.

Within the framework of the German Astrophysical Virtual Observatory (\emph{GAVO}) 
project, the service \emph{TheoSSA} is developed. It allows the \emph{VO} user an easy 
access to synthetic spectral energy distributions (SEDs). We discuss 
quality control problems and the reliability of SEDs provided by \emph{TheoSSA}.
\end{abstract}

\section{Quality Control for Theoretical Data?}
\label{sect:qc}
The question how to introduce quality control for theoretical data is a delicate issue.
Concerning e.g\@. atomic data like oscillator strengths of line transitions,
a comparison with observations will prove their reliability. Different algorithms
for their calculation and parameter regimes for their validity can easily be evaluated.

If complex numerical simulations like model-atmosphere calculations are in focus,
both, 
the code's level of sophistication 
\citep[incl\@. programmed formulae, etc., cf\@. ][]{mihalas_1991, mihalas_2003, rauch_2008}
and 
the atomic-data input 
contribute to the ``quality'' of the code's products -- 
e.g\@. SEDs.

The \emph{VO} provides now a platform where such SEDs of different codes
can be compared to ``cross check'' the codes - data bases like the
T\"ubingen Model-Atom Database
\emph{TMAD} (Sect.~\ref{sect:tmad})
provide model atoms that can be used by every code simply by using an appropriate adapter.

A different kind of quality control is necessary if access to SEDs is provided
via a WWW interface. In case of the
T\"ubingen NLTE Model-Atmosphere Package
(\emph{TMAP}, Sect.~\ref{sect:tmap}),
this is done by 
\emph{TheoSSA} (Sect.~\ref{sect:theossa})
and
\emph{TMAW} (Sect.~\ref{sect:tmaw}).
Since \emph{TMAW} is intentionally restricted in the number of considered
elements and in their model-atom representation, deviations from 
the most elaborated \emph{TMAP} models that consider more elements, atomic levels,
line transitions, etc\@. may occur. In such a case, it appears mandatory that
the \emph{VO} user finds reliable information about the SEDs accuracy
in different wavelength ranges.

\section{A Model-Atmosphere Code: \emph{TMAP}}
\label{sect:tmap}

As an example, we use our
\emph{TMAP}\footnote{http://astro.uni-tuebingen.de/\raisebox{.2em}{\tiny$^\sim$}TMAP}
\citep{werneretal_2003, rauchdeetjen_2003}
that was created in the 1980s and is continuously developed since then. 
With \emph{TMAP}, model atmospheres for hot, compact objects like e.g\@. 
central stars of planetary nebulae, 
PG\,1159 stars,
white dwarfs in novae,
and even neutron stars
can be calculated. Effective temperatures of
20\,000\,K\,$\sla\,T_\mathrm{eff}\,\sla$\,15\,000\,000\,K  
and surface gravities of 
$4\,\sla\,\log g\,\sla\,15$ can be chosen and 
elements from hydrogen to nickel \citep{rauch_2003, rauchetal_2008, rauchetal_2010}
can be included into the calculations. 
\emph{TMAP} considers
hydrostatic and radiative equilibrium,
plane-parallel or spherical geometry,
about 1500 atomic levels treated in NLTE,
about 4000 lines from the elements H - K,
and
presently 200 millions of lines \citep{kurucz_2009}
of Ca - Ni (iron-group elements).

\section{SED Access: \emph{TheoSSA}}
\label{sect:theossa}

At the end of the last century, 
we started to provide 
SEDs of hot,compact stars via our institute's WWW page. 
These were then used e.g\@.
as ionizing spectra in photoionization models of planetary nebulae
because the previously used blackbody fluxes are only a poor
representation of a stellar spectrum. Their flux 
maximum is generally located at lower energies and their peak intensity is
about a factor of three lower \citep[][her Fig\@. 1]{ringat_2012}.

Within a T\"ubingen \emph{GAVO} project, the registered \emph{VO} service
\emph{TheoSSA}\footnote{http://dc.g-vo.org/theossa} was created.
It is a data base to retrieve SEDs of both
pre-calculated model-atmosphere grids and individual models. 
It is presently based on \emph{TMAP} models but
in general prepared to hold SEDs of any
model-atmosphere code.
SEDs that are calculated via \emph{TMAW} (Sect.~\ref{sect:tmaw})
are automatically ingested into \emph{TheoSSA}.
These are identified in
\emph{TheoSSA} by the entry
``DataID.Creator=TMAW'' in the respective meta data.

\section{Individual SEDs on Demand: \emph{TMAW}}
\label{sect:tmaw}

In case that \emph{TheoSSA} contains no SED close enough
to the requested parameter values, the \emph{VO} user is directed to
\emph{TMAW}\footnote{http://astro.uni-tuebingen.de/\raisebox{.2em}{\tiny$^\sim$}TMAW}.
This WWW interface allows to calculate \emph{TMAP} models without
profound knowledge of the code in the background.
In order to keep the calculation time at a reasonable level,
\emph{TMAW}, in contrast to \emph{TMAP},  considers only opacities
of H, He, C, N, and O and uses standard model atoms 
that are provided by \emph{TMAD}. These may be smaller than 
the detailed model atoms that are used for 
highly elaborated, ``final'' models in individual spectral analyses. 

\emph{TMAW} accepts also requests for model grids (in $T_\mathrm{eff}$ and $\log g$).
Up to 150 models are calculated on the PC cluster of the
Institute for Astronomy and Astrophysics in T\"ubingen.
For more extended grids, \emph{TMAW} employs the compute resources of 
AstroGrid-D\footnote{http://www.gac-grid.net}.

\section{A Model-Atom Data Base: \emph{TMAD}}
\label{sect:tmad}

\emph{TMAD}\footnote{http://astro.uni-tuebingen.de/\raisebox{.2em}{\tiny$^\sim$}TMAD}
provides ready-to-use model atoms (in \emph{TMAP} format)
including level energies and radiative and 
collisional transition data. These are continuously updated for the most recent
atomic data.
Presently, it includes the elements 
H, He, C, N, O, F, Ne, Na, Mg, Si, S, Ar, and Ca. Complete model atoms are 
available for model-atmosphere and SED calculations. The latter include fine-structure splitting. 

With suitable adapters, any other model-atmosphere codes can benefit from the data provided by \emph{TMAD}.

\section{A Benchmark Test: The Case of AA\,Doradus}
\label{sect:benchmark}

The initial aim of \emph{TheoSSA} and \emph{TMAW} was to provide easy access to
synthetic stellar fluxes for the PNe community, which was interested in
realistic stellar ionizing fluxes for their PN models (Sect.~\ref{sect:theossa}).

It turned out rather quickly, that other groups were interested in X-ray,
optical, and infrared fluxes as well. Both, \emph{TheoSSA} and \emph{TMAW}, were then
extended for this purpose. While the accuracy in ionizing fluxes is better than
10\,\% compared to \emph{TMAP} calculations with the most detailed model atoms \citep{ringat_2012}, 
the precision in the optical is limited due to the smaller, standard
model atoms that are used in the TMAW model-atmosphere calculations.
The objective was here to perform preliminary spectral analyses with \emph{TMAW} SEDs
and to be better than 20\,\% in $T_\mathrm{eff}$, $\log g$, and
abundance determinations.

A recent spectral analysis of the sdOB primary of the binary AA\,Dor 
\citep{klepprauch_2011}
which is based on optical spectra and highly
elaborated model atmospheres, that 
considered opacities of the elements 
H, He, C, N, O, Mg, Si, P, S, Ca, Sc, Ti, V, Cr, Mn, Fe, Co, and Ni,
provides an ideal test case. 

Therefore, analogously to \citet{klepprauch_2011}, 
we calculated the same model-atmosphere grid with
\Teffw{39\,500 - 43\,500} ($\Delta$\,\Teffw{250}) and 
\loggw{5.30 - 5.60} ($\Delta$\,\loggw{0.01}) 
but with the standard \emph{TMAW} procedure, i.e\@.
only the elements H, He, C, N, and O are considered with the
standard \emph{TMAD} model atoms (Sect.~\ref{sect:tmad}).

We performed a $\chi^2$ test \citep[for further details, see][]{ringat_2012}, 
and determined \Teffw{41\,150} and \loggw{5.43} while 
\citet{klepprauch_2011} found
\Teffw{40\,600} and \loggw{5.46}. 
The deviations are only about 1\,\% in \Teff\ and
7\,\% in \logg .
The example of AA\,Dor shows clearly, 
that \emph{TMAW} models are already well suited for spectral analysis of optical spectra. 

Further improvements of  \emph{TMAW} are currently introduced, 
e.g\@. line-formation calculations with detailed model atoms. In the near future, 
we will extend the considered elements to Ne, Mg, and the so-called iron-group
elements (Ca-Ni).

\newpage

\section{Summary}
\label{sect:summary}

The \emph{VO} constitutes an ideal environment not only to find
theoretical data and simulations but also to compare them -- 
or, in other words, to perform quality control in a broader sense.  
The \emph{VO} user might encounter problems to evaluate the
results of a \emph{VO} query. Thus, it is a challenge for theory 
groups to intensify comparative studies \citep[cf\@.][]{rauch_2012a}.

In case of e.g\@. stellar model atmospheres and SEDs,
a variety of LTE, NLTE, static, and expanding atmosphere codes is available
and used for spectral analysis. 
Their validity ranges are overlapping \citep{rauch_2012b},
allowing to cross check them in detail. 

The \emph{TheoSSA} (Theoretical Stellar Spectra Access) service is designed
to provide SEDs of any kind in a \emph{VO}-compliant form. 
Its efficiency is strongly increasing if more different 
model-atmosphere groups provide their SEDs with a proper description in
their respective meta data.

\vfill

\acknowledgements 
TR is supported by the German Aerospace Center (DLR, grant 05\,OR\,0806),
ER by the Federal Ministry of  Education and Research (BMBF, grant 05A11VTB).
The 
\emph{TheoSSA} 
and
\emph{TMAW}
services 
used to retrieve and calculate theoretical spectra, respectively, 
for this paper were constructed as part of the activities of
the German Astrophysical Virtual Observatory.
We gratefully thank 
AstroGrid-D (http://www.gac-grid.de/) for the computational resources.

\bibliography{P124}

\end{document}